\documentclass[12pt]{article}

\usepackage{epsf}
\usepackage[dvips]{graphicx}
\def\vec#1{\mathbf{#1}}

\begin{document}

\baselineskip 24pt

\begin{center}
{\bfseries INVESTIGATION OF NEUTRON--DEUTERON CHARGE-EXCHANGE
REACTION AT SMALL TRANSFER MOMENTUM}

\vskip 5mm

N. B. Ladygina$^*$

\vskip 5mm

{\small
 {\it
Joint Institute for Nuclear Research, Dubna, Russia
}

}
\end{center}

\vskip 5mm


\noindent
Analysis of the $nd\to p(nn)$ reaction in a GeV-energy region is performed in
 the framework
based on the multiple-scattering theory for the few-nucleon system.
The special kinematic condition, when momentum transfer from neutron beam 
to final
proton closes to zero, is considered. The possibility to extract 
the spin-dependent term of
the elementary $np\to pn $  amplitude from $nd$-breakup process is investigated.
The energy dependence of the ratio
 $R=\frac{d\sigma_{nd}} {d\Omega} / \frac{d\sigma_{np}}{d\Omega}$
is obtained taking account of the final-state interaction of the two outgoing 
neutrons
 in $^1 S_0$-state.

\vspace{1cm}

{\bf PACS:} 21.45.+v, 
11.80.La, 
25.45.-z.

\vspace{10cm}

$^*$ {\it
E-mail: ladygina@sunhe.jinr.ru}

\newpage

{\large\bf 1. Introduction}

The nucleon--deuteron charge-exchange reaction is the subject of the 
investigation in the set of the experiments, which are started
in the VBLHE JINR at STRELA and DELTA SIGMA \cite {delta} setups and 
in COSY \cite {cosy} at
ANKE spectrometer. These experiments are performed in special
kinematics, where transfer momentum from the initial nucleon
to the outgoing fast nucleon is close to zero. The goal of these
experiments is to extract the additional information about
the spin-dependent part of the elementary $np\to pn$ amplitude
from the nucleon--deuteron reaction. First this idea was suggested by 
Pomeranchuk \cite {Pom}
already in 1951. Later, it was shown that in the
plane-wave impulse approximation (PWIA) the differential cross
section and tensor analyzing power $T_{20}$ in the $dp$ charge-exchange
reaction are actually fully determined  by the spin-dependent
part of the elementary $np\to pn$
 amplitudes  \cite {cosy}, \cite {alad}.

However, the relative momentum
of the two slow nucleons is very small under such kinematical conditions,
where momentum of the emitted fast nucleon has the same direction and magnitude
as the beam  (in the deuteron rest frame).
As a consequence,
the final-state interaction (FSI)
effects have to play very important role. 
How much is the FSI influence? It is the general question of our consideration
in this paper. 

We consider the $nd\to pnn$ reaction in kinematics of the
DELTA SIGMA experiment, where the outgoing proton has the same direction as the
projectile neutron and transfer momentum is close to zero. 
The neutron-beam kinetic energy changes from 0.8 up to 1.3 GeV.
The theoretical
approach is based on the Alt--Grassberger--Sandhas
formulation of  the multiple-scattering theory for the three-nucleon system.
We apply the matrix inversion method 
to describe  the FSI contributions. 
Earlier, this formalism has been employed for description of the
deuteron--proton breakup in the GeV region \cite {LSh}- \cite {EPJ}.
All calculations have been performed in the deuteron rest frame. 
The results, which obtained in the present paper, are very interesting, 
from our point of view, and useful for further investigations of this
problem.

The paper is organized as follows. The theoretical formalism
is given in the Section 2 in which  we consider both the general description
of the $nd$ break-up reaction and the special case, where transfer
momentum is close to zero.
 The results are presented in Section 3.
 The ratio of the $nd\to pnn$ differential
cross section 
 to the free $np$ scattering differential cross section at $\theta_{\rm lab} =0$
 has been studied. It is shown that final-state interaction is very
 important, although the FSI contribution decreases with the energy
 increasing. In Section 4 the conclusions are given.

\vspace{1cm}

{\large\bf 2. Theoretical formalism}

In accordance with the Alt--Grassberger--Sandhas (AGS) formalism of 
the three-body collision theory \cite {Alt}- \cite {AGS}, the amplitude of 
the neutron--deuteron charge-exchange reaction,
\begin{eqnarray}
n(\vec p)+d(\vec 0)\to p(\vec p_1)+n(\vec p_2)+n(\vec p_3)~,
\end{eqnarray}
is defined by the matrix element of the transition operator $U_{01}$:
\begin{eqnarray}
\label{first}
{U}_{nd\to pnn} \equiv \sqrt {2} <123|[1-(1,2)-(1,3)] U_{01}|1(23)>=
\delta (\vec p -\vec p_1-\vec p_2 -\vec p_3){\cal J}.
\end{eqnarray}

The state $|1(23)>$ corresponds to the configuration of nucleons 2 and 3 forming the
deuteron state and nucleon 1 being the projectile, whereas the state
$<123|$ represents the free motion of  three nucleons after the reaction.
The permutation operators for two nucleons $(i,j)$ appear in this expression
due to particle identity both in initial and final states.

Iterating the AGS equations up to the second-order terms, we can obtain the expression for 
the transition operator $U_{01}$: 
\begin{eqnarray}
\label{next}
U_{01}=g_0^{-1}+t_2+t_3+t_1 g_0t_2 +t_1 g_0t_3+t_2 g_0t_3+t_3 g_0t_2+O(t^3) .
\end{eqnarray}
 By definition, $t_3=t_{12}$ is a two-nucleon transition operator
(the others can be obtained via  cyclic permutations).
Here, $g_0$ is the free nucleon propagator,
\begin{eqnarray}
\label{g0}
g_0= [E+i0 - K]^{-1},
\nonumber
\end{eqnarray}
with the kinetic-energy operator for the three-body system $K$ and the
on-shell energy $E=E_1+E_2+E_3$.
Obviously, the  term $g_0^{-1}$ does not contribute to
the on-shell amplitude (\ref {first}).

Since we consider only such kinematical conditions, when
outgoing proton carries away significant part of the initial
neutron momentum, we can neglect  the corrections due to the
recoil reaction mechanism, where final fast proton leaves the deuteron
without direct knock-out. In addition, we ignore terms, which correspond to
the double scattering with participation of a fast nucleon.
Then, the matrix element 
$U_{nd \to pnn}$ can be approximated by
\begin{eqnarray}
\label{am}
U_{nd \to pnn}&=&\sqrt {2} <123|[1-(2,3)][1+t_{23}(E-E_1) 
g_{23} (E-E_1)]t_{12}^{sym}|1(23)>,
\end{eqnarray}
where the operator $g_{23} (E-E_1)$ is a free propagator for the
(23)-subsystem and the scattering operator $t_{23}(E-E_1)$
satisfies the Lippmann--Schwinger (LS)
equation with two-body force operator $V_{23}$ as  driving term
\begin{eqnarray}
\label{LS}
t_{23}(E-E_1) = V_{23} + V_{23} g_{23}(E-E_1) t_{23}(E-E_1) .
\end{eqnarray}
 The operator $t_{12}^{\rm sym}$ is symmetrized
NN-operator, $t_{12}^{\rm sym}=[1-(1,2)]t_{12}$.

The expression (\ref {am}) can be schematically presented by the set of graphs
(Fig.1). Here, two first terms correspond to the PWIA and the others represent
the scattering with FSI of the slow nucleons. 
Let us rewrite the matrix element (\ref{am}) indicating explicitly
the particle quantum numbers,
\begin{eqnarray}
\label{to}
U_{nd\to pnn}=\sqrt {2}
<\vec {p_1} m_1 \tau_1,\vec {p_2} m_2 \tau_2,\vec {p_3} m_3 \tau_3|
[1-(2,3)] \omega_{23} t^{\rm sym}_{12} |\vec {p} m \tau ,\psi _{1 M_d 0 0} (23)>,
\end{eqnarray}
where $\omega_{23}=[1+t_{23}(E-E_1) g_{23} (E-E_1)]$ and
the spin and isospin  projections are denoted as
$m$ and $\tau$, respectively. 

In momentum representation the deuteron wave function (DWF) 
 $\psi _{1 M_d}(\vec k)$  with
 spin projection $M_d$ is written as
\begin{eqnarray}
\label{dwfsum}
|\psi _{1 M_d}(\vec k)>=\sum _{L=0,2}\sum_{M_L=-L}^{L}
<L M_L 1 {\cal M}_S|1 M_d> u_L (k) Y_L^{M_L}(\hat k) | 1 M_s>,
\end{eqnarray}
with the spherical harmonics $Y_L^{M_L}(\hat k)$ and the Clebsh--Gordon
coefficients in the standard form.
In our calculations, we use  the pole parameterization of the deuteron 
wave  function \cite {par}:
\begin{eqnarray}
\label{dwf}
u_0 (p)=\sqrt{{2\over\pi}}\sum _{i}\frac {c_i}{\alpha_i ^2 +p^2}~~,~~
u_2 (p)=\sqrt{{2\over\pi}}\sum _{i}\frac {d_i}{\beta_i ^2 +p^2}.
\end{eqnarray}

Inserting in Eq.(\ref {to}) the unity 
\begin{eqnarray}
{\bf 1}=\int d\vec p^\prime
|\vec p^\prime m^\prime\tau ^\prime>
<\vec p^\prime m^\prime\tau ^\prime|,
\nonumber
\end{eqnarray}
and using the definition of the deuteron wave function, Eq.(\ref{dwfsum}),
  we get  the following expression for the reaction amplitude:
\begin{eqnarray}
\label {11}
{\cal J}&=&{1\over 2}
<{1\over 2} m_2 {1\over 2} m_3|S M_S>
 <{1\over 2} m^\prime _2 {1\over 2} m^\prime _3|S M^\prime _S>
<L M_L 1 {\cal M_S} |1 M_d>
 <{1\over 2} m^{\prime\prime }  {1\over 2} m^\prime _3|1 {\cal M _S}>\times
\nonumber\\
&\times &\int d\vec p _0 {^\prime }
\left<\vec p_0, S M_S\left|1 + m_N \frac{t^{ST^\prime =1}
(E-E_1)}
{\vec {p_0} ^2 -\vec {p_0}^{\prime 2}+i0 }\right|
\vec p _0 {^\prime }, S M^\prime _S \right>
u_L (|\vec p_0^\prime -\vec q /2|)Y_L^{M_L}
(\widehat {\vec p_0^\prime -\vec q /2})\times
\nonumber\\
&\times&<\vec {p_1} m_1,(\vec {p_0}^\prime + \vec q/2)~ m^\prime _2|
t_{T=0}^{\rm sym}(E - E_3^\prime)-t_{T=1}^{\rm sym}(E - E_3^\prime)
|\vec p m, (\vec {p_0}^\prime -\vec q/2)~ m^{\prime \prime}>-
\nonumber\\
&~-~&
(2\leftrightarrow 3),
\end{eqnarray}
where  $m_N$ is the nucleon mass, $S$ corresponds to the spin of the
two slow nucleons which participate in the final-state interaction. 
We have 
introduced here the momentum transfer
$\vec q=\vec p -\vec p_1$,  relative momenta
$\vec p_0={1\over 2}(\vec p_2 -\vec p_3)$ and
$\vec {p_0}^\prime ={1\over 2}(\vec {p_2}^\prime  -\vec {p_3}^\prime)$.
Henceforth, all summations over dummy discrete indices are implied.

The two slow neutrons interaction is described by the wave function
\begin{eqnarray}
<\psi^{(-)}_{\vec p_0 S M_S T M_T}|\vec p _0 {^\prime } S M^\prime _S T M_T>
=
\delta (\vec {p_0} - \vec {p_0}^\prime)\delta _{M_s M_s^\prime}+
\nonumber\\
+\frac{m_N}{\vec {p_0} ^2 -\vec {p_0}^{\prime 2}+i0 }
< \vec p_0 S M_S | t^{ST}| \vec p _0 {^\prime } S M^\prime _S >.
\end{eqnarray}
This wave function contains the FSI part, which can be taken into account
  in different ways.

In the present paper we use the matrix inversion method (MIM) suggested in
\cite {HaTa, BJ} and
applied to study the deuteron  electro-disintegration \cite {Sch1,Sch2}
and deuteron proton breakup process \cite {LSh,EPJ}.
As in ref. \cite {Sch1}, we consider the truncated partial-wave expansion,
\begin{eqnarray}
&&<\psi^{(-)}_{\vec p_0 S M_S T M_T}|\vec p_0 {^\prime } S M^\prime _S T M_T>=
\delta _{M_S M^\prime _S} \delta (\vec p_0 -\vec p_0 {^\prime})+
\\
&+&\sum _{J=0}^{J_{\rm max}} \sum _{M_J=-J}^{J}
Y_l^\mu (\hat p_0) <l \mu S M_S|J M_J>
\psi _{l l^\prime} ^\alpha (p_0 {^\prime})
<l^\prime \mu ^\prime S M_S ^\prime|J M_J>
{Y} _{l^\prime}^{\ast \mu ^\prime} (\hat p_0{^\prime}),
\nonumber
\end{eqnarray}
where $J_{\rm max}$ is the maximum value of the total angular momentum in
$nn$ partial waves and $\alpha =\{ J,S,T\}$ is the set of conserved quantum
numbers.

Under kinematical conditions, where
transfer momentum  $\vec q=\vec p -\vec p_1 $ is close to zero, one can anticipate
that the FSI in the $^1S_0$-state is prevalent at comparatively small
$p_0$-values. Then two-neutron wave function is
\begin{eqnarray}
<\psi^{(-)}_{\vec p_0} |\vec p_0 {^\prime }>=
 \delta (\vec p_0 -\vec p_0 {^\prime})+
\frac {1}{4\pi}\psi _{00} ^{001} (p_0 {^\prime}).
\nonumber
\end{eqnarray}
The radial part of this  wave function 
 $\psi _{00} ^{001} (p_0^\prime )$ can be expressed by a series of
$\delta$-functions:
\begin{equation}
\label{psi}
\psi_{00}^{001}(p_0^\prime)=\sum_{j=1}^{N+1}C^{001}(j)\frac {\delta (p_j-p_0)}{p_j^2},
\end{equation}
where $p_j (j=1,\dots ,N)$ are the grid points associated with the Gaussian nodes 
over the interval $[-1,1]$ with dimension equal $N$ and
$p_{N+1}=p_0$.  The coefficients $C(j)$ are determined from the 
solution of the linear algebraic 
equations system approximately equivalent to the Lippmann--Schwinger equation 
for two-neutron scattering
\cite {shebeko}.

In such a way we get the following expression for the amplitude of
the $nd$ charge-exchange process \cite {EPJ}:
\begin{eqnarray}
{\cal J}&=&{\cal J}_{\rm PWIA}+{\cal J }_{^1S_0},
\nonumber\\
\nonumber\\
\label{Jpwia}
{\cal J}_{\rm PWIA}&=&\frac {1}{2} <L M_L 1 {\cal M_S}|1 M_D>\times
\nonumber\\
&\times&\Bigl\{ <{1\over 2} m_2^\prime {1\over 2} m_3|1 {\cal M_S}>
< m_1 m_2,\vec p_1,\vec p_0+\vec q/2|
t^0 -t^1
|\vec p,\vec p_0-\vec q/2, m m_2 ^ \prime >\times
\nonumber\\
&\times&u_L ( |\vec p_0 -\vec q/2| )
Y_L^{M_L}(\widehat { \vec p_0 -\vec q/2})-
\nonumber\\
&-&<{1\over 2} m_2^\prime {1\over 2} m_2|1 {\cal M_S}>
< m_1 m_3,\vec p_1,\vec p_0-\vec q/2|
t^0 -t^1
|\vec p,\vec p_0+\vec q/2, m m_2 ^ \prime >\times
\nonumber\\
&\times&u_L ( |\vec p_0 +\vec q/2| )
Y_L^{M_L}(\widehat { \vec p_0 +\vec q/2})
 \}~,
\\
\nonumber\\
\label{ampl}
{\cal J}_{^1S_0}&=&\frac {(-1)^{1-m_2 -m_2^\prime }}{2\cdot 4\pi} 
\delta _{m_2 ~ -m_3} <L M_L 1 {\cal M_S}|1 M_D>
<{1\over 2} m^{\prime\prime } {1\over 2} -m_2^\prime|1 {\cal M_S}> \times
\\
&\times&\int dp _0 {^\prime } p _0 {^\prime }	^2
< m_1 m _2^\prime ,\vec p_1,\vec p_0^\prime +\vec q/2|
t^0 -t^1
|\vec p,\vec p_0^\prime -\vec q/2, m m ^ {\prime\prime } >
\psi _{00} ^{001} (p_0^\prime )\times
\nonumber\\
&\times& u_L(|\vec p_0^\prime -\vec q/2|)
Y_L^{M_L}(\widehat { \vec p_0^\prime -\vec q/2})~.
\nonumber
\end{eqnarray}

 Since the $q, p_0 \ll p, p_1$,
 we can neglect the $q$- and $p_0$- dependences of the high-energy 
 $np$ $t$-matrix in Eqs.(\ref{Jpwia})-(\ref{ampl}). Besides, the integrand 
 in Eq.(\ref{ampl}) is suppressed at high 
 $p_0^\prime$. It offers us the opportunity to consider $np\to pn$ vertex
 as the free $np$ scattering at  angle $\theta^*=\pi$ in the center-of-mass
 system (c.m.s.) 
 (or $np\to pn$ at $\theta_{\rm lab} =0$).
As well known, the $NN$ $t$-matrix is described in collinear geometry
by the three independent amplitudes:
\begin{equation}
t^{NN}_{\rm cm}(\theta^* =\pi)=A(E) +
[F(E)-B(E)] (\mbox{{\boldmath $\sigma_1$}} \hat p^*) 
(\mbox{{\boldmath $\sigma_2$}} \hat p^*)+
B(E) (\mbox{\boldmath$\sigma_1$}\mbox{\boldmath$\sigma_2$}),
\end{equation}
where $\hat p^*$ is the unit vector in the beam direction in the c.m.s. and $E$ 
is the on-shell energy of two nucleons. 
Note, that this description is correct only for on-energy shell t-matrix
 in the center-of-mass system. 
However, we need the off-energy  shell $t$-matrix in the frame, which in our kinematics
corresponds to  
the laboratory one. In order to relate two these t-matrices, we use some results of the
relativistic potential theory \cite {Hell76, Garc77}. This procedure has
 been presented 
in details in \cite {LSh}. Here we give only the final formula:
\begin{eqnarray}
\label{33}
<m_1 m_2^\prime ,\vec p \vec p_0^\prime|t|
\vec p\vec p_0^\prime,m m^{\prime\prime }>&=&
NN^\prime F
<m_1|D^\dagger (\vec u,\vec p)|\mu_1>\times
\\
&\times&<m_2^\prime|D^\dagger (\vec u,\vec p_0^\prime)|\mu_2^\prime>
 <\mu_1\mu_2^\prime ,\vec p^*|t_{\rm cm}|\mu\mu^{\prime\prime},\vec p^*>
 \times
 \nonumber\\
&\times&<\mu|D (\vec u,\vec p)|m>
<\mu^{\prime\prime }|D(\vec u,\vec p_0^\prime)|m ^{\prime\prime}>,
\nonumber
\end{eqnarray}
where $D$ is a Wigner rotation operator in the spin space and
$u$ is a four-velocity. In our kinematical situation, where
$ p= p_1 \gg  p_0^\prime $, the Wigner rotation operators
can be considered as the unity. In the other words, the spin structure
of the $t_{NN}$-matrix in the reference frame  is the same as
the spin structure of the $t_{NN}$-matrix in the c.m.s. 

The  normalization factors $N$ and $N^\prime$ are determined by the 
Jacobians for the transformations  between the reference frame  and
c.m.s. In our situation these factors are equal to each other: 
\begin{equation}
N=N^\prime =\left [ \frac {m_N+E_p}{4m_NE_p}\sqrt{2m_N(m_N+E_p)}\right ]^{1/2}.
\end{equation}
Other coefficient $F$ is the kinematical factor connected with the transition
from the on-energy shell $t$-matrix to off-energy shell one
\begin{equation}
F=\frac {m_N+E_p}{2E_p}.
\end{equation}
Since we have neglected  the $\vec p_0^\prime$  dependence of
 the high-energy $np$ $t$-matrix,
 Eq.(\ref {ampl}) can be integrated over $\vec p_0^\prime$ taking into account 
Eqs.(\ref {dwf}), (\ref {psi}). In order to simplify final expressions,
we consider here only $S$-component of the deuteron wave function.
This assumption does not to influence on the results, since we do not consider
in the present paper any polarization observables, for which $D$-wave contribution 
is very important.
\begin{eqnarray}
\label{full}
{\cal J}&=&\frac {m_N+E_p}{4E_p}\frac {1}{\sqrt {4\pi}}
\times
\nonumber\\
&\times&\Bigl\{ 
<{1\over 2} m_2^\prime {1\over 2} m_3|1 M_d>
< m_1 m_2,\vec p^* |
t^0_{\rm cm} -t^1_{\rm cm}|\vec p^*, m m_2 ^\prime >u_0(|\vec p_0-\vec q/2|)-
\nonumber\\
&-& <{1\over 2} m_2^\prime {1\over 2} m_2|1 M_d>
< m_1 m_3,\vec p^*|
t^0_{\rm cm} -t^1_{\rm cm}
|\vec p^*, m m_2 ^ \prime >u_0(|\vec p_0+\vec q/2|)+
\\
&+&(-1)^{1-m_2 -m_2^\prime } 
\delta _{m_2, ~ -m_3} 
<{1\over 2} m^{\prime\prime } {1\over 2} -m_2^\prime|1 M_d>\times
\nonumber\\
&\times&\sum _{j=1}^{N+1} \sum_{i}\frac {c_i}{qp_j}Q_0\left(\frac {\alpha_i^2+p_j^2+q^2/4}
{p_jq}\right)C^{001}(j)
< m_1 m _2^\prime ,\vec p^*|
t^0_{\rm cm} -t^1_{\rm cm}
|\vec p^*, m m ^ {\prime\prime } >\Bigr\}.
\nonumber
\end{eqnarray}
Here $Q_0(z)$ is the Legendre polynomials of the second kind.

The cross section of the $nd\to pnn$ reaction is defined in the standard manner:
\begin{equation}
\sigma=(2\pi )^4~ \frac {E_p}{p}\cdot \frac {1}{6} \int ~ d\vec p_1 ~ d\vec p_2 ~
 \delta (M_d+E_p-E_1-E_2-E_3)~~|{\cal J}|^2, 
\end{equation}
where $\vec p_3=\vec p-\vec p_1-\vec p_2$ and 
$E_3=\sqrt {m_N^2+(\vec p-\vec p_1-\vec p_2)^2}$. The squared 
amplitude can be obtained straightforwardly from Eq.(\ref {full}):
\begin{eqnarray}
\label{sqr}
|{\cal J}|^2&=&\frac {1}{2\pi}~\left( \frac {m_N+E_p}{4E_p}\right)^2
\{ (2B^2(E)+F^2(E))[
3u^2_0(|\vec p_0-\vec q/2|)+3u^2_0(|\vec p_0+\vec q/2|)-
\nonumber\\
&-&2u_0(|\vec p_0-\vec q/2|)u_0(|\vec p_0+\vec q/2|)+
\nonumber\\
&+&\left(\sum _{j=1}^{N+1} \sum_{i}\frac {2c_i}{qp_j}Q_0\left(\frac {\alpha_i^2+p_j^2+q^2/4}{p_jq}\right)C^{001}(j)
\right)^2+
\\
&+&\sum_i\frac {2c_i}{qp_j}Q_0\left(\frac {\alpha_i^2+p_j^2+q^2/4}{p_jq}\right)
[C^{001}(j)+C^{001}(j)^*]\times
\nonumber\\
&\times&(u_0(|\vec p_0-\vec q/2|)+u_0(|\vec p_0+\vec q/2|))]+
\nonumber\\
&+&3A^2(E)[u_0(|\vec p_0+\vec q/2|)-u_0(|\vec p_0-\vec q/2|)]^2\}.
\nonumber
\end{eqnarray}
This expression contains both the spin-dependent part of the $np$ amplitude,
 $B$ and $F$ terms, and the spin-independent one, $A$ amplitude.
However, $A$ amplitude is multiplied by the difference of 
the two deuteron wave functions, which depend on the 
practically undistinguished arguments. 
The Eq.(\ref {sqr}) can be significantly simplified by the assumption 
of $\vec q=0$: 
\begin{eqnarray}
|{\cal J}|^2&\approx&\frac {1}{2\pi}~\left( \frac {m_N+E_p}{2E_p}\right)^2 
(2B^2(E)+F^2(E))\left\{
u_0(p_0)+
\sum _{j=1}^{N+1} ~u_0(p_j)C^{001}(j)
\right\}^2~.
\end{eqnarray}
In this expression the term corresponding to the spin-independent part
of the $np\to pn$ amplitude has vanished. Due to that
we get the factorization of the squared $nd\to pnn$ amplitude in the two parts. 
One of them depends on the deuteron and two slow-neutron wave functions.
Other term corresponds to the spin-dependent component of the elementary
$np\to pn$ cross section. This result is very important, since it offers us the
opportunity to extract the spin-dependent part of the neutron--proton 
charge-exchange cross section from more complicated reaction with the deuteron 
participation. But it should be noted, that such factorization is possible
due to special kinematical conditions, when the transfer momentum is small
in respect to the beam one. Moreover, we have used some model to take
account of the two slow neutron final-state interaction. Therefore,
 the obtained result is model-dependent, what does not allow us to 
 correctly extract
the spin dependent part of the $np$ cross section. 

\vspace{1cm}

{\large\bf 3. Results}

In order to relate our calculation with the existing experimental data
\cite {delta}
 we consider the ratio of the $nd$ charge-exchange differential cross section 
to the free $np$ scattering differential cross section at $\theta_{\rm lab} =0$:
\begin{equation}
R=\frac {d\sigma (nd\to pnn)}{dp_1 d\Omega}/\frac {d\sigma (np\to pn)}{d\Omega }.
\end{equation}
The recent energy-dependent phase shift analysis data (PSA) \cite {said}
 have been used for
the determination of the $np$ amplitudes, which are needed to define
 both  the $nd$ and  $np$ charge-exchange
differential cross sections.

 This ratio for the initial neutron kinetic energy $T_n =1~ {\rm GeV}$ is 
 presented
in Fig.2 as a function of the final-proton momentum $p_1$.
 The dashed and solid curves correspond to the PWIA and PWIA+FSI calculations,
 respectively. One can see, the behaviours of these curves are 
significantly distinguished.  
 The solid line  has a very sharp
peak, when the  momentum $p_1$ is close to the beam momentum $p$,
 or transfer momentum $q$ is close to zero, while we do not observe any 
 peak for the dashed line. This peak indicates the FSI contribution to 
the $nd$ differential cross section.
In this region the value of the R ratio varies in 10 times,
 when transfer momentum changes on a few MeV/c. 
Since any experiment has the limited momentum resolution, we consider also the R ratio 
integrated over $p_1$ in some region:
\begin{equation}
 R_{int}=\int_{p-\Delta p}^{p} dp_1 R(p_1)=\int_{p-\Delta p}^{p} dp_1 \frac 
{d\sigma (nd\to pnn)}{dp_1 d\Omega}/\frac {d\sigma (np\to pn)}{d\Omega } ~.
\end{equation}
Here we introduce a new variable $\Delta p$, which is a small difference
between the initial neutron momentum and outgoing proton one.
The integration limits change from
$p-\Delta p$ up to maximal value of $p_1$ equal to $p$. 

 The integrated $R$ ratio  is shown in Fig 3. in dependence on $\Delta p$
 at the neutron kinetic energy $T_n=1 ~{\rm GeV}$. One can see,  the PWIA curve
 is close to the PWIA+FSI one, when $\Delta p$ increases.
 As it follows from Fig.2, the FSI contribution at first increases and
 then decreases the $nd$ differential cross section in respect to the
 PWIA predictions. This influence of the final-state interaction on the 
  $nd$ differential cross section has an effect on the
behaviour of the integrated $R$ ratio.   
In fact, the difference between the PWIA and full calculation 
results is about $30 \%$ 
for $\Delta p$ equal to 10 MeV/c,
about $15 \%$ for $\Delta p$ equal to 20 MeV/c and these  lines 
are practically undistinguished, when $\Delta p$
is equal to 60 MeV/c.

The energy dependence of the integrated $R$ ratio is presented in  Fig.4.
 The integration has been performed
for $\Delta p$ equal to 30 MeV/c. We investigate the energy region between 800 and 
1300 MeV. The upper limit is defined by the existing  phase shift analysis data 
for $np$ scattering. The dash-dotted line is 
   obtained using the result of
  \cite {lbshz} with NN amplitudes taken from  energy-dependent 
 PSA \cite {said}.
The difference between result, obtained taking into account FSI, and PWIA result
 is about 10$\%$ for kinetic energy 800 MeV and few per cent for kinetic energy 1300 MeV.
 Thus, the contribution of the FSI decreases, when the kinetic 
 energy is increasing.
 
\vspace{1cm}

{\large\bf 4. Conclusion}

In this paper the $nd\to p(nn)$ reaction has been studied at the neutron kinetic energy
$T_n=0.8 - 1.3 ~ {\rm GeV}$ in kinematics, where transfer momentum is close to zero.
It was shown that the expression for the $nd\to pnn$ differential
cross section is factorizable in the two parts. One of them is fully
defined by the spin-dependent part of the elementary  $np\to pn$ 
cross section, while the other part depends on the deuteron and
two slow neutrons wave functions. This factorization
allows us to extract
the  spin-dependent part of  the $np$ charge-exchange squared  amplitude, using
the $nd\to p(nn)$ reaction. 
But  obtained result will be dependent on the model, which was applied 
 for FSI description, and the
 choice of the deuteron wave function. This fact does not offer the opportunity 
 to get the precise value of the spin-dependent part of the
  free $np$ scattering amplitude by such method. However, it is possible
to extract some useful information about $np$ charge-exchange process 
(for example, sign, approximate value, etc.).

The other important question, which has been studied, is the role of the
final-state interaction of the two slow neutrons. 
We have considered the $R$ ratio of the $nd$ differential 
 cross section to the
elementary $np\to pn$ one. 
It was shown, that 
the final-state interaction gives a considerable contribution into the three-fold $nd\to pnn$
differential cross section, while the FSI influence on the integrated 
variables is small.
 Moreover, 
 the contribution of 
the FSI into the integrated $R$ ratio decreases with the increasing energy.

\vspace{1cm}
The author is thankful to Drs. V.P. Ladygin, F. Lehar and V.I. Sharov for fruitful discussions.
This work has been supported by the Russian Foundation for Basic Research
under grant  $N^{\underline 0}$  07-02-00102a.

\newpage
\begin{center}
{\large\bf Figure captions}
\end{center}

\vspace {1cm}

{\bf Fig.1.} Graphic representation of the amplitude of the $nd\to pnn$
 reaction.
 
{\bf Fig.2.} $R$ ratio  vs. the fast proton momentum $p_1$ at $T_n=1 ~{\rm GeV}$.

{\bf Fig.3.} Integrated $R$ ratio as a function of $\Delta  p$ at 
$T_n=1 ~ {\rm GeV}$.

{\bf Fig.4.} Energy dependence of the integrated $R$ ratio.

\newpage
\begin{figure}[h]
 \epsfysize=150mm
 \centerline{
  \epsfbox{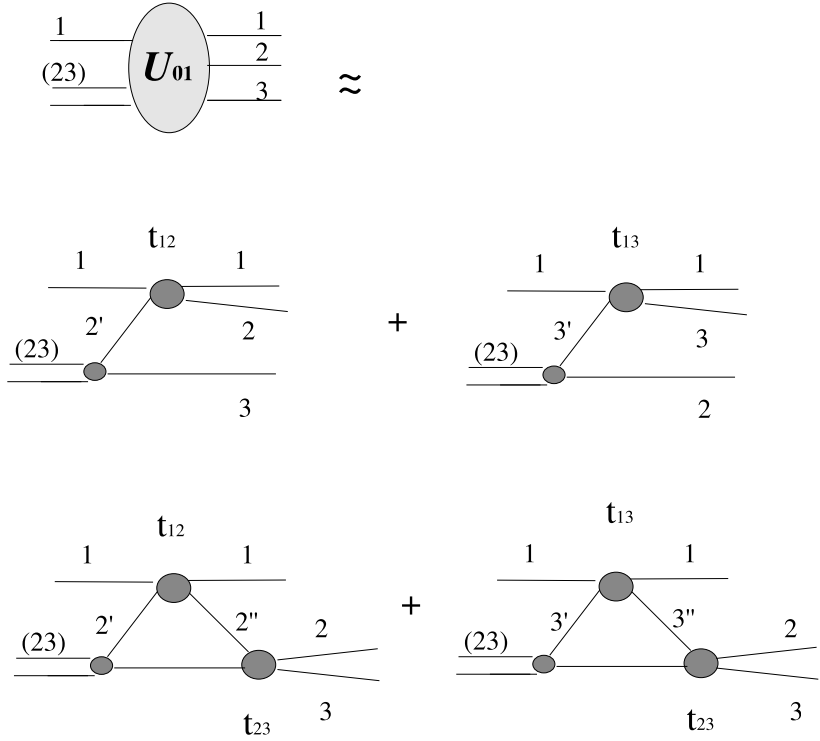}}
 \caption{Graphic representation of the amplitude of the $nd\to pnn$
 reaction. }
\end{figure}

\newpage
\begin{figure}[h]
 \epsfysize=120mm
 \centerline{
 \epsfbox{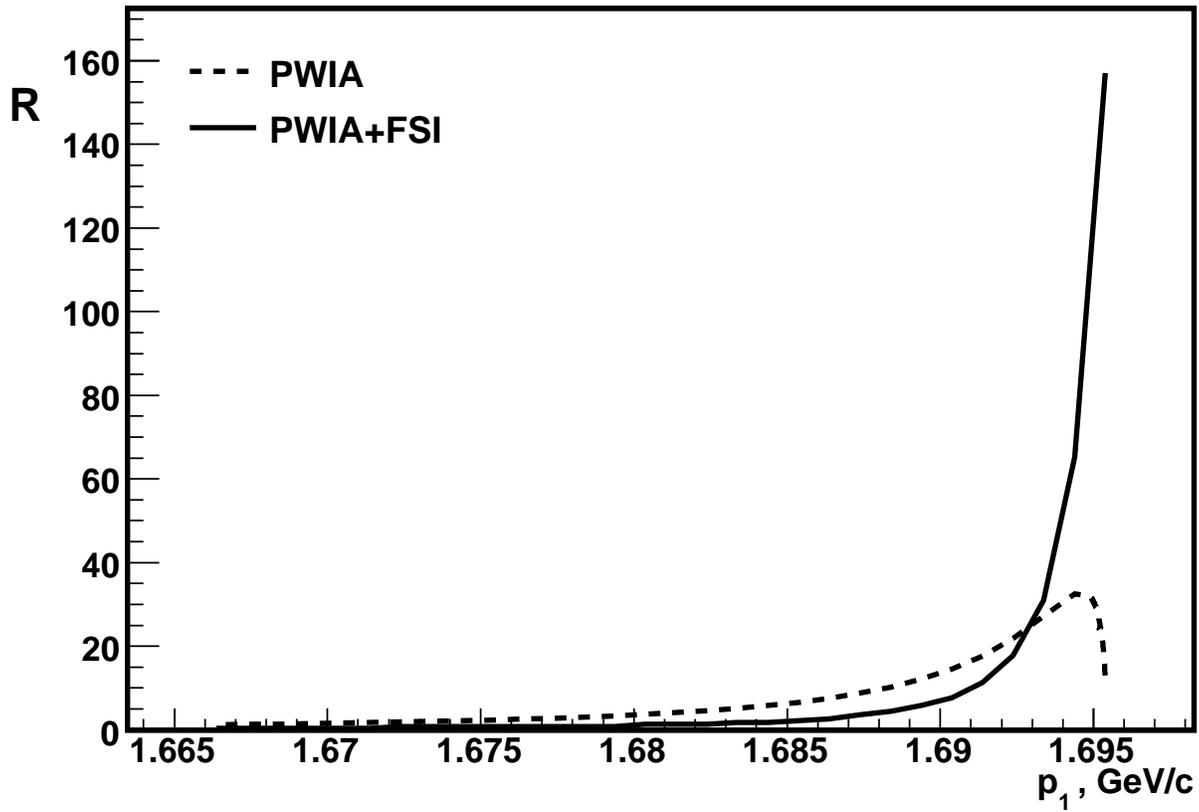}}
 \caption{$R$ ratio vs. the fast proton momentum $p_1$ at $T_n=1 ~{\rm GeV}$.}
\end{figure}

\newpage
\begin{figure}[h]
 \epsfysize=120mm
 \centerline{
 \epsfbox{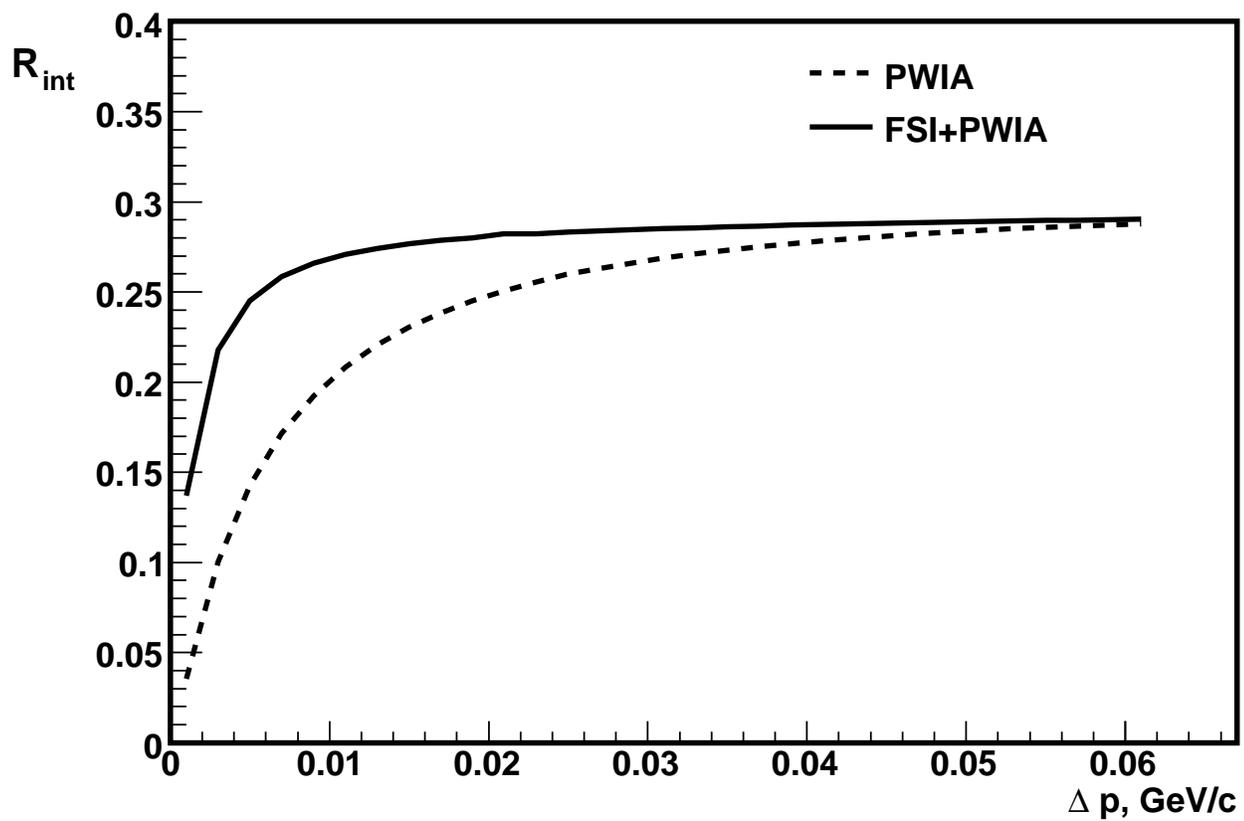}}
 \caption{Integrated R ratio as a function of $\Delta  p$ at $T_n=1 ~ {\rm GeV}$. }
\end{figure}

\newpage
\begin{figure}[h]
 \epsfysize=120mm
 \centerline{
 \epsfbox{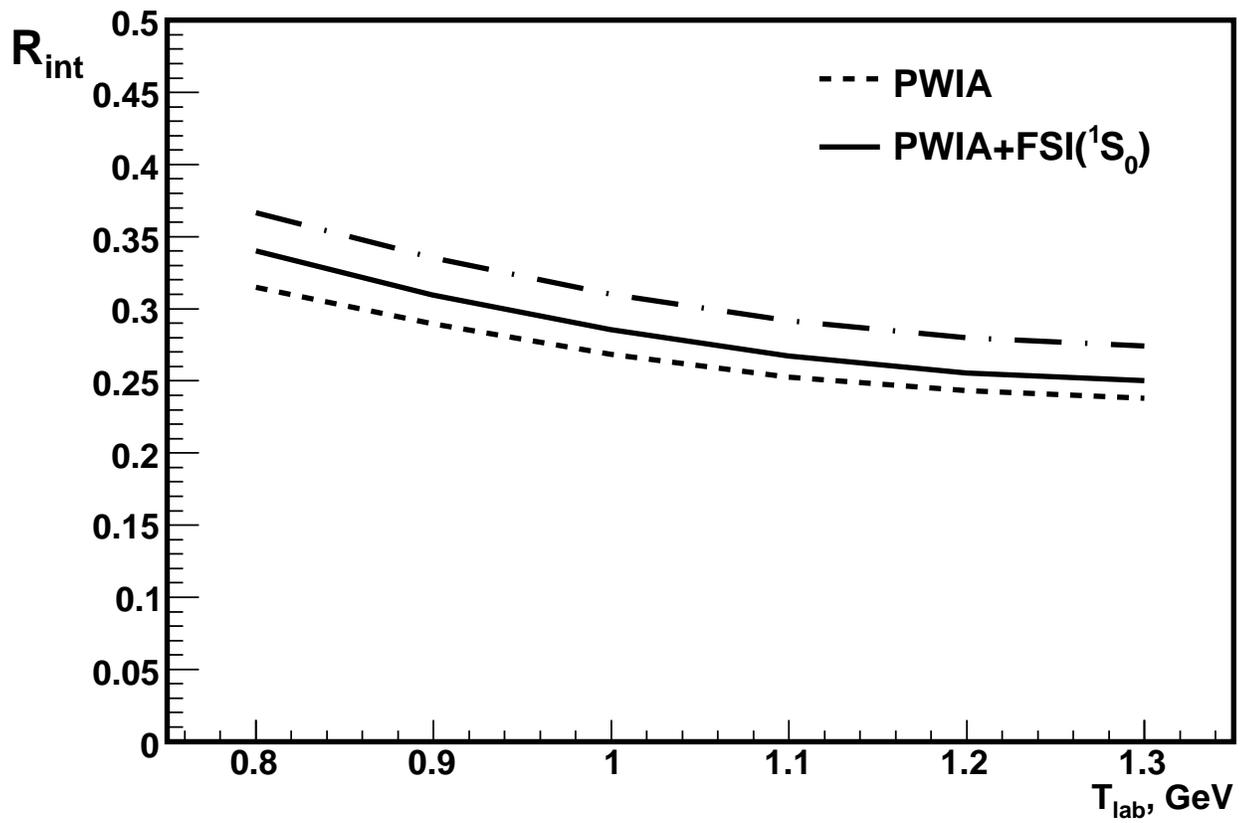}}
 \caption{Energy dependence of the integrated R ratio.}
\end{figure}

\end{document}